\documentclass[aps,preprint,showpacs]{revtex4}
\usepackage{epsfig}
\usepackage{latexsym}
\begin{document}
\draft

\title{Giant Planar Hall Effect in Epitaxial (Ga,Mn)As Devices}
\author{H.X. Tang$^1$, R.K. Kawakami$^2$, D.D. Awschalom$^2$, and M.L. Roukes$^1$}
\address{$^1$Condensed Matter Physics 114-36, California Institute of
Technology, Pasadena, CA 91125, USA\\
$^2$Department of Physics, University of California, Santa Barbara, CA 93106, USA
}
\date{April 4, 2002}

\begin{abstract}
\vspace{1in}
Large Hall resistance jumps are observed in microdevices patterned from epitaxial 
(Ga,Mn)As layers when subjected to a swept, in-plane magnetic field. This giant 
planar Hall effect is four orders of magnitude greater than previously observed 
in metallic ferromagnets. This enables extremely sensitive measurements of the 
angle-dependent magnetic properties of (Ga,Mn)As. The magnetic anisotropy fields
deduced from these measurements are compared with theoretical predictions.
\end{abstract}
\maketitle

Ferromagnetic semiconductors are of considerable current 
interest since they offer prospects for realizing
semiconducting spintronics devices that have no analogs in metallic 
ferromagnetic system.~\cite{citation1,citation2} 
One recent and striking example is the electric 
field control of ferromagnetism.~\cite{citation3} Semiconductor-based magnetic materials also offer 
new possibilities for attaining great improvements in performance over metallic 
magnetic devices. Among examples here is conductivity-matching to attain efficient spin injection 
into semiconductors.~\cite{citation4}

Semiconductor ferromagnetism also gives rise to new physical phenomena because it
is possible to engineer, and enhance, spin-orbit coupling in ways that are not
possible in metallic systems. In this letter we report the observation of a 
giant planar ``Hall'' effect (GPHE)  
in epitaxial (Ga,Mn)As thin film devices. This is manifested as a spontaneous transverse
voltage that develops, because of spin-orbit copling, in response to longitudinal
current flow in the absence of an applied field. Analogous effects studied in metallic ferromagnets,
are exceedingly small –- typically of order m$\Omega$.~\cite{citation6} Related phenomena
have recently been investigated in ferromagnetic semiconductors ~\cite{citation7}, but here we
report the previously unrecognized, and quite remarkable, response of the GPHE to an applied 
in-plane magnetic field. Within the high quality, single domain ferromagnetic semiconductors 
investigated here we find a switchable effect that is about four orders of magnitude stronger
than found in metallic ferromagnets. Below  
we describe measurements that take advantage of this strong GPHE to provide insight, and
unprecedented resolution, into the mechanism of magnetic switching
within these materials. These data, in turn, enable complete characterization of the magnetic anisotropy
of the (Ga,Mn)As epilayers. At present, the large magnitude of this PHE is not fully understood.  
We presume this phenomenon stems from the combined effects of significant spin-orbit coupling 
in the valence band of the zincblende crystal structure, and the large spin polarization of holes 
in (Ga,Mn)As. The temperature dependence of the magnetization and the coercivity determined 
by electrical measurement should provide additional insight into the underlying physical mechanisms.    

Molecular beam epitaxy (MBE) at 250$^{\circ}$C was employed to deposit 150nm-thick 
Ga$_{0.948}$Mn$_{0.052}$As epilayers on top of an insulating GaAs(001) substrate with 
a buffer layer. Various thicknesses and concentrations of materials have been 
investigated, and some samples are annealed at elevated temperatures.  
The devices described herein are patterned from a single wafer (UCSB–-001115A, 
Curie temperature $T_c\sim 45$K), however, it is notable that all devices studied 
(fabricated from a variety of different epilayers) exhibit consistent behavior.  

Magnetoresistance measurements are carried out on families of Hall bars 
(widths 6\hspace{3pt}$\mu$m--1\hspace{3pt}mm), and square van der Pauw 
devices (3$\times$3\hspace{3pt}mm$^2$). The former are aligned along the [110] 
direction by a combination of photo- and electron-beam- lithography. 
Voltage probes on the Hall bars are carefully designed to minimize their 
perturbation upon current flow within the devices (cf. SEM micrograph, Fig.~1e). 
Standard four-probe lock-in measurements are performed by a 10\hspace{3pt}nA 
ac sensing current at 14\hspace{3pt}Hz,  
excitation is intentionally kept quite low to obviate electron heating.  
Magnetic fields are generated using a 3-coil superconducting magnet that allows 3D field 
orientation without physically disturbing the sample.  We execute two 
classes of experiments, the first in which the field orientation is fixed 
in-plane along a specific direction, $\varphi_H$, with respect to the
longitudinal axis of the Hall bars, while the field magnitude is swept linearly 
between $\pm$1000\hspace{3pt}Oe.  In the second, we fix the magnitude of the 
applied in-plane field, while stepping its orientation clockwise or counter-clockwise.  
Prior to each sweep, an in-plane field of 6000\hspace{3pt}Oe is applied to saturate 
the sample magnetization, $\mathbf{M}$. 

At all angles, except for those along (110) directions, two 
abrupt jumps are observed in PHE measurements.  Families of 
data taken from Hall bars spanning from 6\hspace{3pt}$\mu$m to macroscopic 
(1\hspace{3pt}mm) dimensions are shown in Fig.~1(a-c). These are obtained for 
orientation 20$^{\circ}$ away from [110].  For comparison, the field-dependent 
sheet resistance of a 100\hspace{3pt}$\mu$m Hall bar is also displayed in Fig.~1d.

Four distinct features emerge.  a) Large switching events at distinct 
magnetic fields are observed in the Hall resistance; these are accompanied 
by small jumps (relative to the background) in the longitudinal resistance.  
b) Between these switching fields, the planar Hall resistance remains constant 
at approximately 37 $\Omega$. The signal polarity reverses at each switching event.  
c) The switching fields appear to be independent of sample size and geometry.  
Measurements on samples with square, van der Pauw geometry, as large as 
$3 {\times} 3\hspace{3pt}$mm$^2$, exhibit identical switching behavior as those of the smaller, 
micron-scale devices –- even though the magnitude of the Hall resistance 
is reduced in the former, presumably due to non-uniform current distribution.  
d) When the width of the Hall device is reduced to $\sim$6\hspace{3pt}$\mu$m, 
small Barkhausen jumps are observed.  These occur in close proximity 
to the switching transitions (Fig.~1f), and appear to demonstrate that 
the propagation of domain walls is constrained by geometry.~\cite{citation8}

We have also investigated samples with Hall bars fabricated along many 
other directions besides the [110] crystalline axis.  
We find that the switching fields do not depend on the orientation of 
the Hall bars, whereas the magnitude of planar Hall resistance jumps 
are systematically reduced as one moves away from the (110) directions. 

Figure~2a presents the dependence of $R$--$H$ loops upon field orientation angle 
$\varphi_H$  as it is varied from $-30^{\circ}$ to 30$^{\circ}$ in the plane. In the field 
range of these experiments, only one jump occurs along the (110) 
directions. Away from these special orientations, a two-jump reversal 
is always observed. The first switching field $H_{c1}$  is almost constant, 
while the second switching field $H_{c2}$ decreases dramatically and approaches
$H_{c1}$ at around $\pm$30$^{\circ}$. 

We interpret the jumps in the Hall resistance as follows. The electric 
field within a single domain ferromagnetic film with in-plane magnetization 
can be written as~\cite{citation5}
\begin{eqnarray}
\label{eq1a}
E_{x}& = & j\rho _{\perp }+j(\rho _{\parallel}-\rho _{\perp })\cos ^{2}\varphi \\
\label{eq1b}
E_{y} & = &j(\rho _{\parallel}-\rho _{\perp })\sin \varphi \cos \varphi
\end{eqnarray}
where the current density $j$ is assumed to be uniformly distributed along 
the Hall bar, $x$ and $y$ are the longitudinal and transverse axes, 
and $\varphi$ is the angle between the magnetization and 
current density $j$. $\rho _{\parallel}$ and $\rho _{\perp}$  
are the resistivities for current oriented parallel and perpendicular 
to the magnetization. The anisotropic magnetoresistance phenomenon is 
described by Eq.~\ref{eq1a}.  The transverse resistance, \textit{i.e.} the planar Hall 
resistance, is expressed in Eq.~\ref{eq1b}; it exhibits extrema at $\varphi$ = 45$^{\circ}$ 
and its cubic equivalents.  To verify this angular dependence of the 
planar Hall resistance, an in-plane field of constant magnitude 6000\hspace{3pt}Oe is 
applied to saturate the magnetization, and its orientation is swept 
through 360 degrees (Fig.~2c).  In accordance with Eq.~\ref{eq1b}, the measured 
Hall resistance exhibits extrema for applied field orientations of 
$\sim$45$^{\circ}$, $\sim$135$^{\circ}$, $\sim$225$^{\circ}$, and $\sim$315$^{\circ}$. Note that the first maximum 
of planar Hall resistance appears at 135$^{\circ}$ instead of 45$^{\circ}$, indicating 
$\rho_{\parallel}-\rho_{\perp}<0$ ($-73\hspace{3pt}\Omega$ from Eq.~\ref{eq1b}). 
This property of (Ga,Mn)As is distinct from that in conventional 
ferromagnetic metals, where $\rho_{\parallel}-\rho_{\perp}>0$. 
It may originate from the different manner in which holes and electrons 
contribute to the spin-orbit interaction in ferromagnetic materials.

The anomalous switching behavior of the Hall resistance shown in Figs.~1(a-c) 
can be explained by a two-jump sequence of magnetization: [100] 
($\varphi \sim$ -45$^{\circ}$) $\rightarrow$ [010] ($\varphi \sim$ 45$^{\circ}$)$\rightarrow$
[$\bar{1}$00] ($\varphi \sim$ 135$^{\circ}$).  This evolution also accounts for the 
accompanying small longitudinal resistance jumps shown in Fig.~1d.  Between the switching events, 
the sample remains in what appears to be a \textit{macroscopic} single domain state.  
In this regime the magnetization evidently rotates coherently according to 
Stoner-Wohlfarth model,~\cite{citation9} hence the planar Hall resistance continues 
to evolve to a small degree with field. 
Scanning SQUID microscopy of (Ga,Mn)As epilayers magnetized in-plane have provided evidence for the existence 
of macroscopic single domains on length scales extending to hundreds of microns.~\cite{citation11}
Domain states within the sample exist only in the vicinity of the 
switching field, and the associated domain wall scattering evidently generates the 
remarkable resistance spikes shown in Fig.~1d.

Figure~2b summarizes the signatures of the coercive fields manifested in our 
electrical transport measurements. The field 
loci delineating the resistance transitions are shown in polar coordinates.  
The $H_{c1}$ lines form a 
rectangular shape, whereas the $H_{c2}$ lines are more complicated.  
The latter follow the extrapolation of $H_{c1}$ lines at low field but 
evolve towards the (110) axes in higher fields.  Eventually, at a 
field around 2500\hspace{3pt}Oe, the second jump becomes smeared and  
reversible.  These measurements clearly elucidate behavior that is generic in our (Ga,Mn)As epilayers:
the in-plane magnetic anisotropy is nearly cubic, but it is biased by a small two-fold preference along 
[110].

Unusual multiple switching, somewhat analogous to that demonstrated in this 
work, has also been observed in ultrathin epitaxial Fe films, through 
the magneto-optic Kerr effect (MOKE).  A switching pattern analogous to that 
of Fig.~2b was measured in Ag/Fe/Ag(001) system by Cowburn \textit{et al.},~\cite{citation12}
although with significantly less resolution in their metallic system. 
To explain their results a simple model is invoked, incorporating a 
well-defined domain wall pinning energy into a complex, anisotropic 
magnetocrystalline energy surface (A weak in-plane uniaxial anisotropy is 
superimposed along one easy axis of a strong cubic anisotropy).  Our 
experimental data can be explained via similar 
domain reversal energetics, but in the present case the in-plane uniaxial easy 
axis is collinear with a \textit{hard} axis of the cubic anisotropy.  The 
corresponding free energy density of such a single domain magnet can be written as, 
$E=K_u \sin^2 {\varphi} + ({K_1}/{4}) \cos^2 {2\varphi} - M H \cos(\varphi - {\varphi}_H)$. 
Here $K_u$, $K_1$ are in-plane uniaxial and cubic anisotropy constants. The equilibrium state is defined by the conditions, 
${\partial  E}/{\partial \varphi} = 0$ and ${{\partial ^2} E}/{\partial {\varphi}^2} > 0$. The former gives,
\begin{equation}
\label{eq3}
K_u \sin 2 \varphi - K_1 \sin 4 \varphi +
M H \sin(\varphi - {\varphi}_H) = 0. 
\end{equation}
At zero field, four 
distinct magnetization states, corresponding to four local energy minima, 
can exist: ${\varphi}_{1,2} ^0=\pm (\pi/4-\delta)$,${\varphi}_{3,4} ^0=\pm (3\pi/4+\delta)$, 
with $\delta=\sin ^{-1}(K_u / K_1)$. 
Domains can exist over short length scales 
in a demagnetized thin film. Upon application of an in-plane field, 
these small-scale domains quickly become suppressed and the whole sample evolves 
into a macroscopic single domain state with $\varphi$ close to one of the zero fields minima $\varphi^{0}_{1-4}$.  
When the external field is reversed, 
magnetization reversal is achieved via an intermediate state corresponding to the sample 
magnetization oriented almost orthogonally (90$^{\circ}$) to the initial and final 
directions of the magnetization.  Domain states mediate the transitions 
from one energy minimum to another.  For a domain wall to become 
liberated to propagate through the sample, the reversed external field must be 
increased to the point where a characteristic pinning energy density, $\epsilon$, 
is exceeded, \textit{i.e.},
$\mathbf{H} _c \cdot (\mathbf{M} _2 - \mathbf{M} _1 ) = \epsilon$.
Here $\mathbf{M}_1, \mathbf{M}_2$ are the initial and final magnetization, and 
$\mathbf{H}_c$ is the switching field. 
If $\mathbf{H}_c$ is small compared to the 
cubic anisotropy field, coherent rotation of $\mathbf{M}_1$ and $\mathbf{M}_2$ 
from the zero-field equilibrium is negligible.  
For transitions from [100] to [010], a $\sim$90$^{\circ}$ domain wall with core 
magnetization along [110] is required to propagate across the sample, 
giving $\mathbf{H}_c \cdot \hat{x} = -\epsilon _{110} / 2 M
\sin(45^{\circ}-\delta)$, in which $\epsilon_{110}$  is the corresponding 
domain wall pinning energy density.  Considering all possible orientational 
trajectories, we can describe the loci of transitions as 
$\mathbf{H} \cdot \hat{x} = \pm \epsilon _{110} /2 M \sin(45^{\circ} -\delta)$ and 
$\mathbf{H} \cdot \hat{y} = \pm \epsilon _{\bar{1}10} /2 M \sin(45^{\circ} +\delta)$.
At low field, these describe two parallel sets of lines that are in 
excellent correspondence with the switching pattern observed in our 
experiments (Fig.~2b).  At high fields two new pieces of physics 
become important.  First, coherent rotation of \textbf{M} must be 
considered and, second, the energy density of a domain wall 
also becomes significantly reduced.~\cite{citation13}
As a result, high field transitions progressively evolve towards 
the (110) directions. 

Several additional points are worthy of mention.  First, PHE measurements 
enable determination of crystallographic 
orientation with remarkable precision; we estimate that the angular error 
in establishing the (110) crystalline axes is less than 0.04$^{\circ}$.  Second, 
apart from the singularity along these (110) directions, neither single
 transitions nor three-transition processes are observed.  This justifies 
our assumption that in-plane uniaxial anisotropy does \textit{not} exist 
along the cubic easy axes.~\cite{citation14}  

We are able to deduce both the cubic and uniaxial anisotropy fields 
through PHE measurements.  To achieve this, a large, constant magnetic
field is applied in the plane while its orientation is rotated continuously.  
Fig.~2c shows data from such measurements for clockwise and 
counterclockwise sweeps of magnetic field orientation, for a 
family of field magnitudes.  
When $H<H_{cA}$, where $H_{cA}\sim$ 2500\hspace{3pt}Oe is the dominant cubic anisotropy 
field, the magnetoresistance reverses each time the magnetization switches 
across the cubic hard axis.  The planar Hall resistance becomes 
reversible for fields greater than 2500\hspace{3pt}Oe, in which case the magnetization 
rotates coherently according to Stoner-Wohlfarth model. Thus, for a given external field angle 
$\varphi _{H}$, the macroscopic in-plane magnetization orientation, $\varphi$, 
can be calculated by using expression in Eq.~\ref{eq1b}. Fitting all the 
computed data sets ($\varphi_{H} ,\varphi$) to Eq.~\ref{eq3}, we consistently and unambiguously 
extract the anisotropy fields  $H_{cA} = 2K_1/M = 2400$\hspace{3pt}Oe,  
$H_{uA} = 2K_u/M = 160$\hspace{3pt}Oe.   

Progress has recently been made toward gaining 
a theoretical understanding of magnetic anisotropy in III-V magnetic 
semiconductors.~\cite{citation15,citation16}   It is generally 
agreed that, in addition to an intrinsic cubic anisotropy, 
(Ga,Mn)As possesses a substantial out-of-plane uniaxial component 
with sign that is dependent on whether tensile or compressive biaxial 
strain exists at the interface.~\cite{citation7} 
While attention has focused almost exclusively on the out-of-plane magnetic anisotropy,
recent work on (Ga,Mn)As magnetic tunnel junctions ~\cite{TanakaHigo} highlights the importance of 
the in-plane anisotropy. 
On the other hand, theoretical models of cubic anisotropy predict 
that the in-plane cubic easy axes can be oriented along either (100) 
or (110) depending upon the hole concentration and the degree of 
spin-splitting.    
By contrast, all of our data to date, on a variety of (Ga,Mn)As epilayers, 
indicate that the cubic easy axes are aligned along (100) axes.  
Further PHE studies on additional epilayers are needed to determine 
if the cubic anisotropy exhibits the variations predicted by theory.  
Our experimental results do agree, however, with predicted magnitude of the 
cubic anisotropy field, which we find to be about 2400\hspace{3pt}Oe. 

We have also studied the temperature dependence of PHE, which should 
be of significant importance in elucidating its underlying physical 
mechanisms.  Fig.~3a shows the results for 
a 10\hspace{3pt}$\mu$m Hall bar, measured under conditions of careful temperature 
regulation, stepped downward from 50K to 0.32K, for fixed-orientation, 
swept-magnitude, applied magnetic fields.  The magnitude of both  
PHE and the coercive fields increases rapidly with decreasing temperature 
(Fig.~3b). For $T < 10$K, both the PHE and sheet resistivity appear to diverge 
logarithmically down to the lowest measured temperatures, while the ratio 
$\Delta R _H / R _{\Box}$ remains nearly constant.  Here, 
$\Delta R _H$ is the PHE resistance jump and  $R _{\Box}$ is 
the zero-field sheet resistance.  
This ratio, $\Delta R _H / R _{\Box}$, should provide valuable information about the hole spin polarization.  
We find that it decreases monotonically 
with increasing temperature, qualitatively tracking the magnetization 
of a 3$\times$3\hspace{3pt}mm$^2$ sample measured by SQUID magnetometry(Fig.~3c).  

In conclusion, these first observations of a giant planar Hall effect in (Ga,Mn)As devices enable 
systematic investigation of in-plane magnetic anisotropy and magnetization 
reversal via electrical transport measurements. In semiconducting materials, GPHE measurements provide unique
advantages over magneto-optical techniques. For example, carrier concentration
changes arising from sample illumination can be circumvented.  
It is also notable that the high signal-to-noise attainable in GPHE measurements 
permits observation of behavior that emerges only in structures of reduced dimensions
(\textit{e.g.} Barkhausen jumps in 6\hspace{3pt}$\mu$m devices). Given the minimal excitation power
required, this technique is compatible with very low temperature 
measurements ($\sim$mK), thus offering new possibilities 
for investigations in micro- and nanoscale spintronic devices.

We gratefully acknowledge support from DARPA under grants DSO/SPINS–-MDA 
972-01-1-0024 (Caltech) and DARPA/ONR N00014-99-1-1096 (UCSB), and from 
the AFOSR under grant F49620-02-10036 (UCSB).   We also thank 
Prof. P.E. Wigen for valuable discussions.

\newpage
\begin{figure}
\begin{center}
\epsfig{file=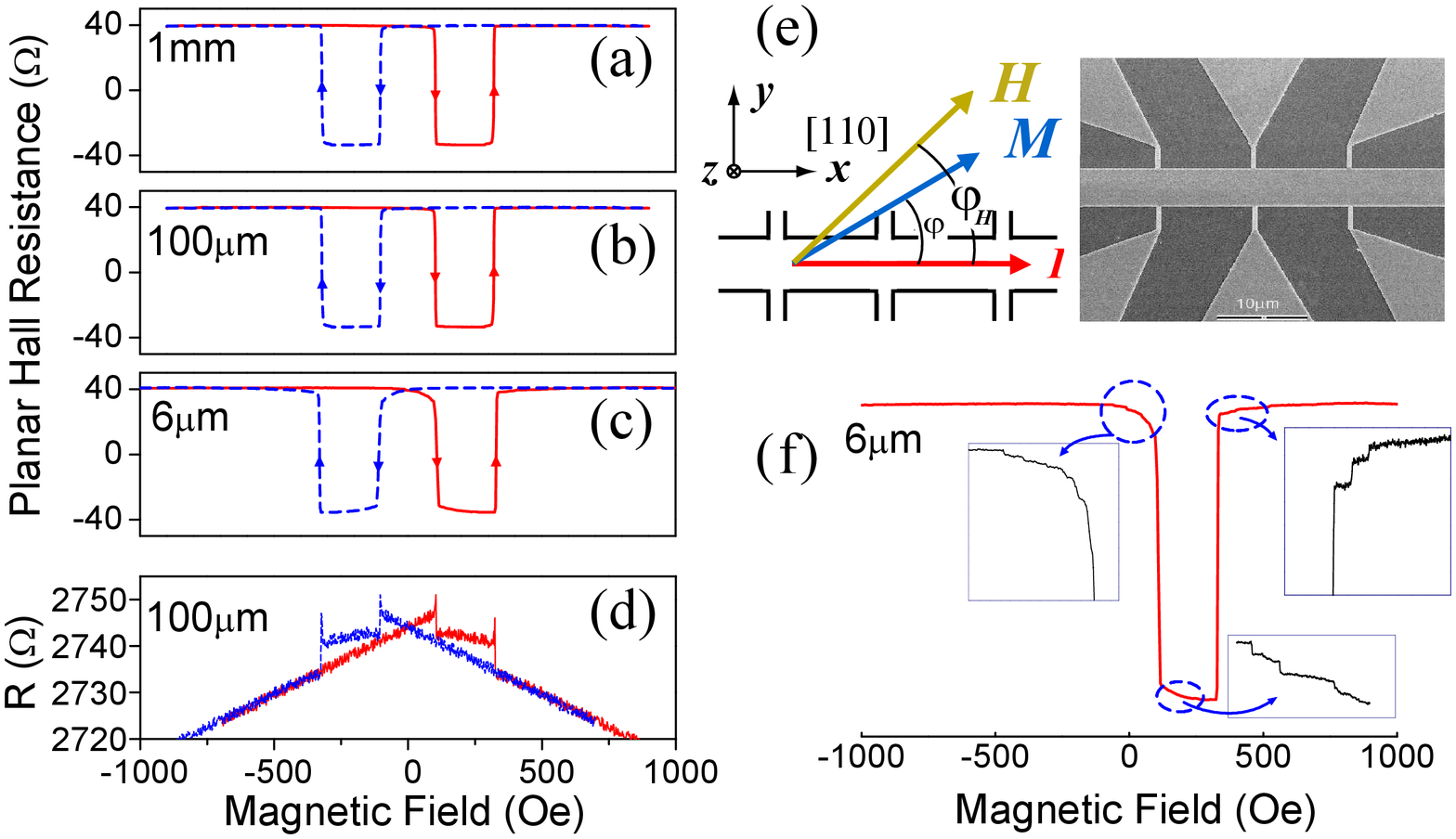,width=6in}
\end{center}
\caption{(a)-(c) Planar Hall resistance for Hall bars (1\hspace{3pt}mm, 100\hspace{3pt}$\mu$m,
6\hspace{3pt}$\mu$m wide) at 4.2K as a function of in-plane magnetic field (at fixed orientation
$\varphi _H = 20^{\circ}$). (d)  Field-dependent sheet resistance of 
a 100\hspace{3pt}$\mu$m-wide Hall bar.  (e) Sketch of the relative orientations of 
sensing current $I$, external field $\mathbf{H}$ and magnetization $\mathbf{M}$.  
A SEM micrograph of a 6\hspace{3pt}$\mu$m-wide device is also shown.  
(f) Barkhausen jumps that are evident solely in 6\hspace{3pt}$\mu$m-wide devices 
near the resistance transitions.}
\label{figure1}
\end{figure}

\newpage
\begin{figure}
\begin{center}
\epsfig{file=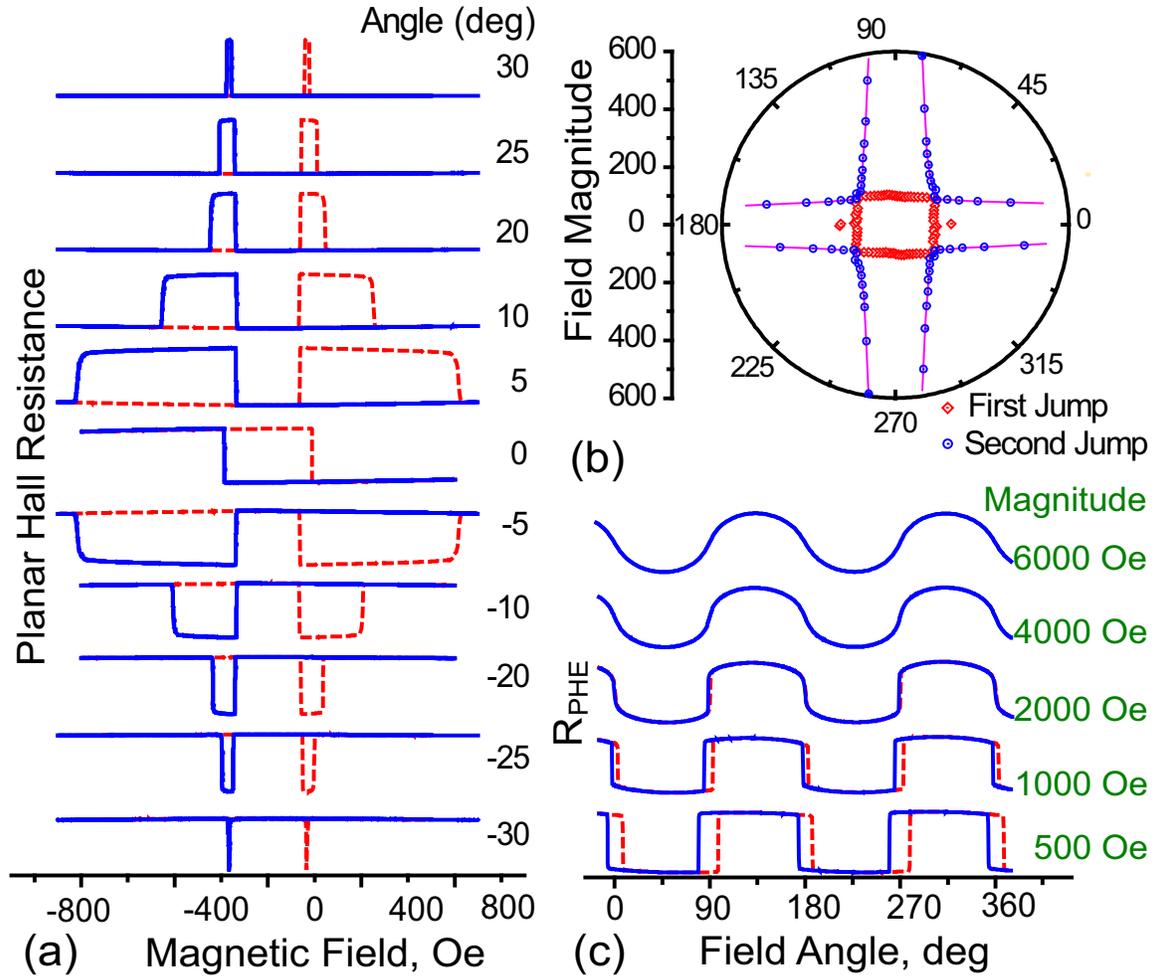,width=6in}
\end{center}
\caption{(a) Angular dependence of the PHE.  
The data are generated by fixed-orientation, swept-field method. (see text)
(b) Polar plot of the first and second switching fields vs. field orientations.
(c) PHE for a family of fixed-magnitude 
sweeps of magnetic field orientation. }
\label{figure2}
\end{figure}

\newpage
\begin{figure}
\begin{center}
\epsfig{file=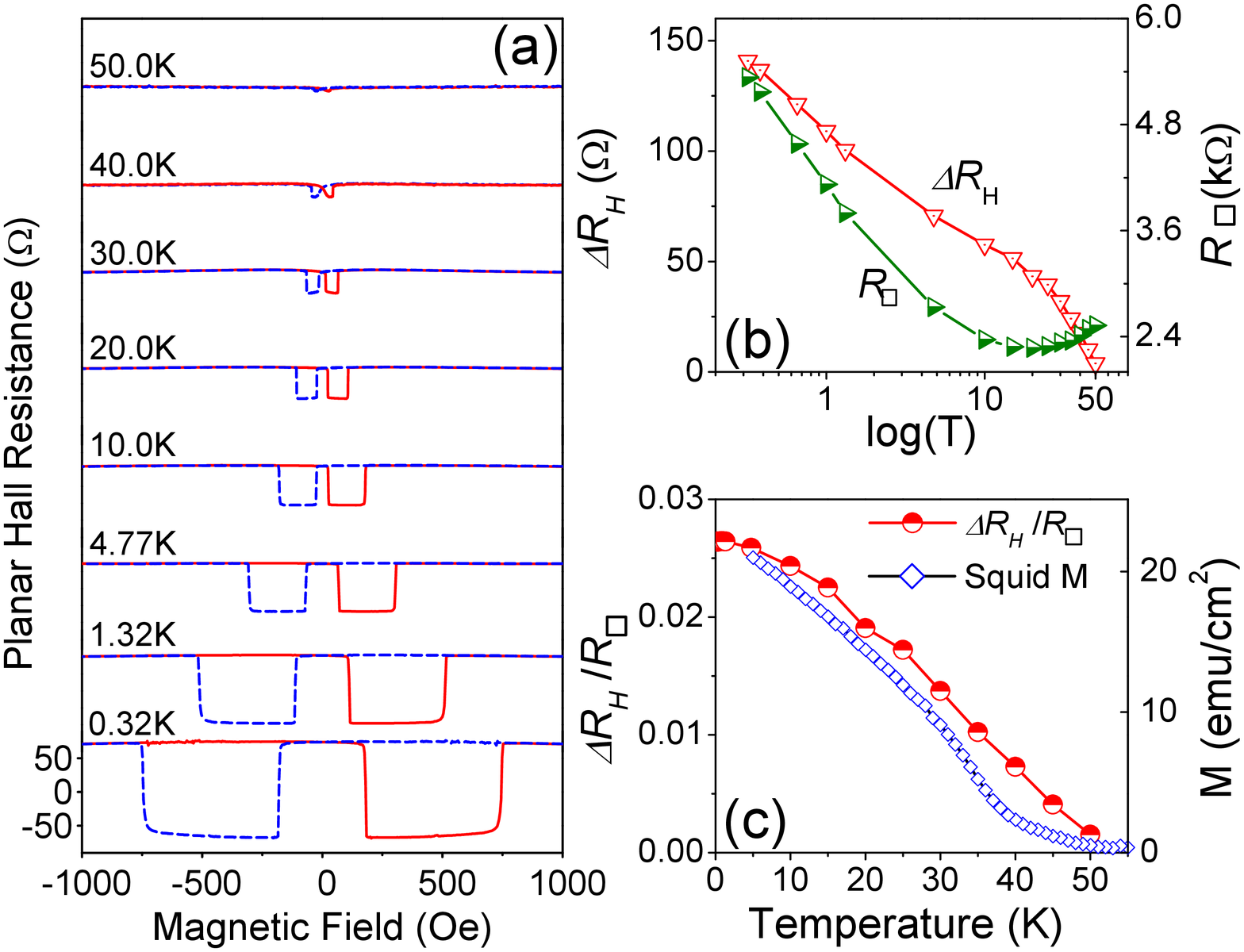,width=6in}
\end{center}
\caption{(a) Temperature dependence of the PHE.  
(b) Semilog plots of the planar Hall resistance jump, $\Delta R _H$ , 
and sheet resistance, $R _{\Box}$, \textit{vs} temperature.  
(c) Comparison between the ratio, $\Delta R _H / R _{\Box}$, 
measured on a 10\hspace{3pt}$\mu$m wide Hall device, and the sample magnetization, 
$M$, measured by SQUID magnetometry on a macroscopic 
(3 ${\times}$ 3\hspace{3pt}mm$^2$) sample. }
\label{figure3}
\end{figure}

\end{document}